\documentclass[prd,preprint,nofootinbib,tightenlines,showpacs]{revtex4}
\usepackage{graphicx}
\usepackage{amsfonts}
\usepackage{latexsym}
\usepackage{amssymb}
\usepackage{epsfig}

\newcommand{\CO}{{\cal O}}

\newcommand{\bear}{\begin{array}}  \newcommand{\eear}{\end{array}}
\newcommand{\bea}{\begin{eqnarray}}  \newcommand{\eea}{\end{eqnarray}}
\newcommand{\beq}{\begin{equation}}  \newcommand{\eeq}{\end{equation}}
\newcommand{\bef}{\begin{figure}}  \newcommand{\eef}{\end{figure}}
\newcommand{\bec}{\begin{center}}  \newcommand{\eec}{\end{center}}
  
\newcommand{\lmk}{\left(}  \newcommand{\rmk}{\right)}

\newcommand{\bib}{\bibitem}

\def\APJ#1#2#3{Astrophys. J. {\bf #1}, #2 (19#3)}
\def\APJJ#1#2#3{Astrophys. J. {\bf #1}, #2 (20#3)}

\def\JHEPP#1#2#3{J. High Energy Phys. {\bf #1}, #2 (20#3)}

\def\MNRAS#1#2#3{Mon. Not. R. Astron. Soc. {\bf #1}, #2 (19#3)}
\def\MNRASS#1#2#3{Mon. Not. R. Astron. Soc. {\bf #1}, #2 (20#3)}

\def\NAT#1#2#3{Nature (London) {\bf #1}, #2 (19#3)}

\def\NPB#1#2#3{Nucl. Phys. {\bf B#1}, #2 (19#3)}
\def\NPBB#1#2#3{Nucl. Phys. {\bf B#1}, #2 (20#3)}

\def\PLB#1#2#3{Phys. Lett. B {\bf #1}, #2 (19#3)}

\def\PLBold#1#2#3{Phys. Lett. {\bf#1B}, #2 (19#3)}

\def\PRD#1#2#3{Phys. Rev. D {\bf #1}, #2 (19#3)}
\def\PRDD#1#2#3{Phys. Rev. D {\bf #1}, #2 (20#3)}

\def\PRL#1#2#3{Phys. Rev. Lett. {\bf#1}, #2 (19#3)}
\def\PRLL#1#2#3{Phys. Rev. Lett. {\bf#1}, #2 (20#3)}

\begin{document}

\title{Axino warm dark matter and $\Omega_b - \Omega_{DM}$ coincidence}

\author{Osamu Seto}
\affiliation{
 Instituto de F\'{i}sica Te\'{o}rica,
 Universidad Aut\'{o}noma de Madrid,
 Cantoblanco, 28049 Madrid, Spain
}

\author{Masahide Yamaguchi}
\affiliation{ 
 Department of Physics and Mathematics,
 Aoyama Gakuin University,
 Sagamihara 229-8558, Japan
}

%
\begin{abstract}
We show that axinos, which are dominantly generated by the decay of 
the next-to-lightest supersymmetric particles produced from 
the leptonic $Q$-ball ($L$-ball), become warm dark matter suitable
for the solution of the missing satellite problem and the cusp
problem. In addition, $\Omega_b - \Omega_{DM}$ coincidence is naturally
explained in this scenario.
\end{abstract}

\preprint{IFT-UAM/CSIC-07-15} 

\vspace*{3cm}
\maketitle


\section{Introduction}

Recent observations of cosmic microwave background anisotropies such as
Wilkinson Microwave Anisotropy Probe measured the abundance of
components of the Universe very precisely. However, their origins are
still one of the major mysteries of cosmology and particle physics. The
fact that the abundances of dark matter and baryon are of the same order
may give us a great hint for their origins.

In the minimal supersymmetric standard model (MSSM), flat directions
consist of squarks/sleptons and produce a non-zero baryon or lepton
number through the Affleck-Dine (AD) mechanism \cite{AD}. Then,
$Q$-balls, which are non-topological solitons \cite{coleman}, can be
produced due to the instability and absorb almost all the produced
baryon or lepton numbers \cite{KasuyaKawasaki}. In the gravity mediated
supersymmetry breaking model, the mass per charge of a $Q$-ball is
larger than that of a nucleon so that they are unstable against the
decay into light fermions. Then, they can directly decay into baryons and the
lightest supersymmetric particle (LSP). In case that the charge of the
produced $Q$-balls is large enough, $Q$-balls can survive even after the
freeze-out of weakly interacting massive particles (WIMPs). 
Thus, the reason why energy densities of dark
matter and baryon are almost the same magnitude can be explained in this
scenario \cite{KS,QballEnqvist}.

However, it was pointed out that LSPs are often overproduced by the
decay of $Q$-balls if the LSP is the lightest neutralino in the MSSM
\cite{EM2}, which gives the stringent conditions on the neutralino LSPs
and AD fields. Only a few models free from this overproduction have been 
proposed~\cite{FHY,FH}.
Instead, the supergravity models in which the LSPs is a
stable gravitino are investigated \cite{Seto}. In this scenario,
$Q$-balls decay into the next-to-lightest supersymmetric particle (NLSP)
directly instead of the LSP gravitino. The LSP gravitinos are produced by 
the decay of NLSP and becomes dominant over other gravitinos produced by
thermal processed \cite{BBP,Pradler} and by the decay of thermal produced NLSP
\cite{FRT,FST} in case that $Q$-balls can survive the evaporation.
Then, it is found that such a gravitino dark matter scenario is still
viable if the late decay of NLSP does not spoil the success of Big Bang
Nucleosynthesis (BBN).  Another interesting possibility is that the LSP
is an axino, which is the fermionic superpartner of an axion. Axinos are
also produced by the decay of NLSPs produced from the $Q$-ball
\cite{RoszSeto} as well as by thermal processes and by the decay of thermally 
produced NLSP \cite{RTW,KMN,CKR,ckkr,Brandenburg}.

Such gravitinos and axinos often become an ideal candidate for cold dark
matter. The models of cold dark matter (CDM) and dark energy combined
with inflation-based scale-invariant primordial density fluctuations
have succeeded at explaining many properties of the observed universe,
especially the large scale structure of the universe. However, going
into the smaller scales, some observations on galactic and subgalactic
($\lesssim$ Mpc) seem to conflict with predictions by high-resolution
N-body simulations as well as analytic calculations based on the
standard CDM model. The first discrepancy is called the missing
satellite problem \cite{MSP}. The CDM-based models predict an order of
magnitude higher number of halos than those actually observed within the
Local Group. The other is called the cusp problem \cite{CP}. The
CDM-based models also predict overly cuspy mass profile for the CDM
halos compared to actual observations within the Local Group. In order
to reconcile such discrepancies, several authors proposed modifications
to the standard CDM-based model though the photoionization mechanism may
overcome such difficulties \cite{BKW}. One method is to reduce the
small-scale power of primordial density fluctuations, which can be
realized in a specific model of inflation~\cite{Yokoyama}. 
Another is to change the
properties of dark matter. Spergel and Steinhardt introduced strong
self-interaction among cold dark matter particles (collisional CDM),
which enhances satellite destruction and suppress cusp formation
\cite{SS}. The warm dark matter \cite{WDM}, which can have relatively
large velocity dispersion at the epoch of the matter-radiation equality,
can also reduce satellite production and cusp formation.

In this paper, we consider axinos dominantly generated by the decay of
NLSPs produced from the leptonic $Q$-ball ($L$-ball). 
Such axinos become warm
dark matter suitable for the solution of the missing satellite problem
and the cusp problem. In addition, $\Omega_b - \Omega_{DM}$ coincidence
is naturally explained through the Affleck-Dine mechanism and the
subsequent $L$-ball formation in this scenario. In the next section, we
discuss $\Omega_b - \Omega_{DM}$ coincidence based on the Affleck-Dine
mechanism and the subsequent $L$-ball formation. In section III, we show
that axinos in our scenario become warm dark matter suitable for the
solution of the missing satellite problem and the cusp problem. In the
final section, we give concluding remarks.

\section{$\Omega_b - \Omega_{DM}$ coincidence from Affleck-Dine leptogenesis
}

We now discuss baryogenesis via Affleck-Dine leptogenesis and dark
matter production from $Q$-ball decays within the framework of gravity
mediated supersymmetry breaking.

\subsection{Lepton asymmetry}

The potential of the AD flat direction field is, in general, lifted by
soft supersymmetric (SUSY) breaking terms and non-renormalizable
terms~\cite{Ng,DineRandallThomas}.  The full potential of the AD field
is given by
\begin{eqnarray}
V(\phi) &=&
\left(m_{\phi}^2\left[1+K\ln\left(\frac{|\phi|^2}{\Lambda^2}\right)\right]
- c_1 H^2 \right)|\phi|^2  \nonumber \\ &&
+\left[\left(c_2 H+ Am_{3/2}\right)\lambda\frac{\phi^n}{nM^{n-3}}+
{\rm H.c. } \right] +\lambda^2\frac{|\phi|^{2n-2}}{M^{2n-6}}.
\label{ADPotential}
\end{eqnarray}
Here, $m_{\phi}$ is the soft SUSY breaking scalar mass for the AD field
with radiative correction $K\ln|\phi|^2$. A flat direction dependent
constant, $K$, takes values from $-0.01$ to $-0.1$~\cite{K}. $\Lambda$
denotes a renormalization scale and $-c_1 H^2$ represents the negative
mass squared induced by the SUSY breaking effect which comes from the
energy density of the inflaton, with an order unity coefficient $c_1 >
0$~\cite{DineRandallThomas}. $\lambda$ is the coupling of a
nonrenormalizable term and $M$ is some large scale acting as its
cut-off. Terms proportional to $A$ and $c_2$ are the A-terms coming from
the low energy SUSY breaking and the inflaton-induced SUSY breaking, 
respectively, where $m_{3/2}$ denotes the gravitino mass.  Here, we
omitted possible terms which may appear by thermal
effects~\cite{ThermalMass,TwoLoop}. These terms are negligible as long
as we consider a sufficient low reheating temperature after inflation,
as we will.  Moreover, the model would face with ``gravitino
problem''~\cite{GravitinoProblem}, if the reheating temperature after
inflation is so high that these thermal effect become effective, unless
gravitino is LSP~\cite{Seto}.

The charge number density for the AD field $\phi$ is given by $n_q=
iq(\dot{\phi}^*\phi-\phi^*\dot{\phi})$ where $q$ is the baryonic (or
leptonic) charge for the AD field. By use of the equation of motion of
the AD field, the charge density can be rewritten as
\begin{equation}
n_q(t) \simeq \frac{1}{a(t)^3}\int^t dt' a(t')^3 \frac{2q\lambda m_{3/2}}{M^{n-3}}
{\rm Im} (A \phi^n) ,
\end{equation}
with $a(t)$ being the scale factor. When the AD field starts to
oscillate around the origin, the charge number density is induced by the
relative phase between A-terms. By taking into account $s =
4\pi^2g_*T^3/90 $, the charge to entropy ratio after reheating is
estimated as
\begin{eqnarray}
\frac{n_q}{s}
 = \left.\frac{T_R n_q}{4M_P^2H^2}\right|_{t_{\rm{os}}}
 \simeq  \frac{q|A|\lambda m_{3/2}}{2}\frac{T_R |\phi_{\rm{os}}|^n}
 {M_P^2H_{\rm{os}}^3 M^{n-3}} \sin\delta.
\label{b-sRatio}
\end{eqnarray}
Here, $M_P \simeq 2.4 \times 10^{18}$ GeV is the reduced Planck mass,
$t_{\rm{os}}$ is the time of the start of the oscillation and
$\sin\delta$ is the effective $CP$ phase. In case that thermal corrections
are ineffective, $H_{\rm os} \simeq m_{\phi}$, which yields
\begin{equation}
|\phi_{\rm os}| \simeq \left(\frac{m_{\phi} M^{n-3}}{\lambda}\right)^{1/(n-2)}.
\end{equation}

From now on, as a concrete example, we consider a $L L \bar{e}$
direction of $n=6$ as the AD field for our scenario. Since this is a
pure leptonic direction, the lepton asymmetry generated by the
Affleck-Dine mechanism can be estimated as
\begin{eqnarray}
\frac{n_L}{s}
\simeq 1\times 10^{-10} \frac{q|A|\sin\delta}{\lambda^{1/2}}
 \left(\frac{m_{3/2}}{100\rm{GeV}}\right)
 \left(\frac{10^3 \rm{GeV}}{m_{\phi}}\right)^{3/2}
 \left(\frac{T_R}{100 \rm{GeV}}\right)\left(\frac{M}{M_P}\right)^{3/2} .
\label{BaryonAsymNoEarly}
\end{eqnarray}

\subsection{Baryon asymmetry and LSP production from $Q$-balls}

The produced lepton asymmetry is not directly released to thermal
bath. Instead, $L$-balls are formed due to the instability and almost
all produced lepton numbers are absorbed into $L$-balls
\cite{KasuyaKawasaki}. 

First of all, we briefly summarize relevant properties of $Q$-balls in
gravity mediated SUSY breaking models. The radius of a $Q$-ball, $R$, is
estimated as $R^2 \simeq 2/(|K|m_{\phi}^2)$~\cite{QballEnqvist}.
Numerical calculations provide a fitting formula for the $Q$-ball charge
\begin{equation}
Q \simeq \bar{\beta}\left(\frac{|\phi_{\rm{os}}|}{m_{\phi}}\right)^2 \times
\left\{
\begin{array}{ll}
\epsilon  \;\quad {\rm for} \quad \epsilon \gtrsim \epsilon_c \\
\epsilon_c  \quad {\rm for} \quad \epsilon < \epsilon_c
\end{array}
\quad ,
\right.
\end{equation}
 with
\begin{equation}
\epsilon \equiv \left.\frac{n_L}{n_{\phi}}\right|_{t_{\rm{os}}} \simeq
 2q|A| \frac{m_{3/2}}{m_{\phi}}\sin\delta 
\end{equation}
 where $ \epsilon_c \simeq 10^{-2} $ 
 and $\bar{\beta}= 6\times 10^{-3}$~\cite{KasuyaKawasaki}.
The $Q$-ball charge can be evaluated as
\begin{eqnarray}
Q \sim 2 \times 10^{20}
\left(\frac{\epsilon}{4 \times 10^{-1}}\right)
\left(\frac{1 {\rm TeV}}{m_{\phi}}\right)^{3/2}
\left(\frac{M^3}{\lambda M_P^3}\right)^{1/2},
\label{QballCharge}
\end{eqnarray}
where we assumed $\epsilon > \epsilon_c$ because it looks to be more
natural than the other which can be realized only 
for an accidental small $\sin\delta$ .  
Furthermore, if $\epsilon < \epsilon_c$,
additional ``unnatural'' parameters are required for our scenario, as we
will show.

A part of the charge of a $Q$-ball can evaporate by the interaction with
particles in the thermal bath. The evaporation of charge of $Q$-ball is
done by the evaporation with the rate
\begin{eqnarray}
\Gamma_{{\rm evap}} \equiv \frac{d Q}{d t} 
 = -4 \pi R_{\rm Q} D_{\rm ev} n^{\rm eq}
 \simeq -4 \pi R_{\rm Q}^2 D_{\rm ev} \mu_{\rm Q}T^2, 
 \label{evaporation}
\end{eqnarray}
with $D_{\rm ev} \lesssim 1$ and by the diffusion with the rate
\begin{eqnarray}
\Gamma_{{\rm diff}} \equiv \frac{d Q}{d t} =
 -4 \pi k R_{\rm Q} D_{\rm diff} n^{\rm eq}
 \simeq -4 \pi k R_{\rm Q} D_{\rm diff} \mu_{\rm Q}T^2, 
 \label{diffusion}
\end{eqnarray}
where $\mu_{\rm Q}$ is the chemical potential of $Q$-balls and the
numerical constant $k$ is very close to unity so that we will drop it
hereafter. $D_{\rm diff} \approx a/T$ is a diffusion
constant~\cite{BanerjeeJedamzik} and $a$ is a particle dependent
coefficient given by~\cite{Joyce:1994zn,DW}
\begin{equation}
a \simeq 
\left\{
\begin{array}{ll}
  4 \qquad {\rm for \quad squark} \\
  6 \qquad {\rm for \quad quark}  \\
  100 \qquad {\rm for \quad left-handed \,\, (s)lepton} \\
  380 \qquad {\rm for \quad right-handed \,\, (s)lepton} \\
\end{array}
\quad .
\right.
\end{equation}
Here, we see that both the evaporation and the diffusion are efficient
for low temperature, from Eqs.~(\ref{evaporation}) and (\ref{diffusion})
with the relation between the cosmic time and the temperature:
\begin{equation}
\frac{dt}{dT} = 
\left\{
\begin{array}{ll}
 \frac{-8}{\pi}\sqrt{\frac{10}{g_*}}\frac{T_R^2 M_P}{T^5} \quad {\rm for} \quad 
T \gtrsim T_R \\
 \frac{-3}{\pi}\sqrt{\frac{10}{g_*}}\frac{M_P}{T^3}  \qquad {\rm for} \quad T < 
T_R
\end{array}
\quad .
\right.
\end{equation}
Moreover, by comparing $\Gamma_{{\rm diff}}$ and $\Gamma_{{\rm evap}}$,
\begin{eqnarray}
\frac{\Gamma_{{\rm diff}}}{
\Gamma_{{\rm evap}}} 
 \simeq \left(\frac{m_{\phi}}{T} a \sqrt{\frac{|K|}{2}}\right)
 \left(\frac{1}{D_{\rm ev}}\right) , 
\end{eqnarray}
we can find that for low temperature
\begin{eqnarray}
 T \lesssim  a \sqrt{\frac{|K|}{2}} m_{\phi} 
 \sim 10 m_{\phi}
 \left(\frac{a}{10^2}\right)\left(\frac{|K|}{10^{-2}}\right)^{1/2},
\end{eqnarray}
the diffusion is more crucial for estimation of the evaporated charge
from $Q$-ball. Equation (\ref{diffusion}) is rewritten as
\begin{eqnarray}
\frac{d Q}{d T} \simeq -4 \pi R_{\rm Q} D_{\rm diff} \mu_{\rm Q}T^2 
 \left(\frac{dt}{dT}\right). 
 \label{diffusion2}
\end{eqnarray}
Integrating Eq.~(\ref{diffusion2}) from $m_{\phi}$, because the
evaporation from $Q$-ball is suppressed by the Boltzmann factor, we can
estimate the total evaporated charge as
\begin{eqnarray}
\Delta Q &\simeq& 32 k R_{\rm Q} \sqrt{\frac{10}{g_*}} a \mu_{\rm Q}
 \frac{T_R^2 M_P}{3 m_{\phi}^3} \nonumber \\ 
 & \sim & \frac{3.2 \times 2.4}{\sqrt{10}} \times 10^{19} 
  \left(\frac{1 {\rm TeV}}{m_{\phi}}\right)^3
  \left(\frac{T_R}{1 {\rm TeV}}\right)^2
 \left(\frac{a}{300}\right)\sqrt{\frac{200}{g_*}}\sqrt{\frac{0.01}{|K|}}
 \left(\frac{\mu_{\rm Q}}{m_{\phi}}\right)
\end{eqnarray}
for $m_{\phi} \gtrsim T_R $,
\begin{eqnarray}
\Delta Q \sim \frac{3.6 \times 2.4}{\sqrt{10}} \times 10^{19} 
 \left(\frac{1 {\rm TeV}}{m_{\phi}}\right)
 \left(\frac{a}{300}\right)\sqrt{\frac{200}{g_*}}
 \sqrt{\frac{0.01}{|K|}}\left(\frac{\mu_{\rm Q}}{m_{\phi}}\right) 
\end{eqnarray}
for $m_{\phi} < T_R$. By taking Eq.~(\ref{QballCharge}) into account, we
obtain
\begin{eqnarray}
\frac{\Delta Q}{Q} \sim \frac{3}{2} \times 10^{-1} 
 \left(\frac{4 \times 10^{-1}}{\epsilon}\right)
 \left(\frac{a}{300}\right)
 \left(\frac{m_{\phi}}{1 {\rm TeV}}\right)^{1/2}
 \sqrt{\frac{\lambda M_P^3}{M^3}}
 \sqrt{\frac{200}{g_*}}\sqrt{\frac{0.01}{|K|}}
 \left(\frac{\mu_{\rm Q}}{m_{\phi}}\right)
\label{FracEvapCharge}
\end{eqnarray}
for $m_{\phi} < T_R$ \footnote{For $m_{\phi} \gtrsim T_R $, the result
is of the same magnitude but with different dependence on $m_{\phi}$
and $T_R$.} and find that about $10 \%$ of $Q$-ball charge would be
evaporated. Here, one can see why the case of $\epsilon < \epsilon_c$ is
irrelevant for us. If $\epsilon < \epsilon_c \simeq 10^{-2}$, $\epsilon$
is replaced with $\epsilon_c$ in Eq.~(\ref{FracEvapCharge}). Then these
$Q$-balls cannot survive the evaporation unless the AD field mass is
extremely small as $m_{\phi}= \cal{O}$$(10)$ GeV or
$(\lambda^{1/3}M_P/M)^{3/2} \ll 1$ .

The evaporated charges are released into the thermal bath so that a part
of them is transformed into baryonic charges through the sphaleron
effects \cite{sphaleron}. Then, the resultant baryon asymmetry is given
as
\begin{eqnarray}
\frac{n_b}{s} &=& \frac{8}{23} \times 
 \frac{\Delta Q}{Q} \times \frac{n_L}{s} \nonumber \\
 &\simeq& \frac{8}{23} \Delta Q\times 10^{-30} 
 \left(\frac{10^3 \rm{GeV}}{m_{\phi}}\right)^{-1}
 \left(\frac{T_R}{100 \rm{GeV}}\right) \nonumber \\
 &\simeq& 10^{-10} \left(\frac{a}{300}\right)\sqrt{\frac{200}{g_*}}
 \sqrt{\frac{0.01}{|K|}}\left(\frac{\mu_{\rm Q}}{m_{\phi}}\right)
 \times
\left\{
\begin{array}{ll}
 \left(\frac{1 \rm{TeV}}{m_{\phi}}\right)^2\left(\frac{T_R}{1 
\rm{TeV}}\right)^3 \quad {\rm for} \, m_{\phi} \gtrsim T_R \\
 \left(\frac{T_R}{1 \rm{TeV}}\right) \qquad \qquad \quad \; {\rm for} \, 
m_{\phi} < T_R
\end{array}
.
\right.
\label{BaryonAsymmetry} 
\end{eqnarray}
Interestingly, the baryon asymmetry does not depend on the effective $CP$
phase $\sin\delta$ unlike usual Affleck-Dine baryogenesis, because the
$CP$ phase dependences in both lepton asymmetry $n_L/s$ and the charge of
$Q$-ball $Q$ cancel each other.  In addition, for $m_{\phi} < T_R$, the
baryon asymmetry basically depends on only one free parameter, the
reheating temperature $T_R$, because other parameters are not free but
known in a sense.

When $Q$-balls decay, the supersymmetric particles are released from them.
Since the $Q$-ball consists of scalar leptons, the number of the produced
supersymmetric particles is given by
\begin{eqnarray}
Y_{\rm NLSP} &=& N_Q \frac{n_L}{s} \nonumber  \\
 &=& 2\times 10^{-9}
 \left(\frac{N_Q}{1}\right)
\left(\frac{\epsilon}{4 \times 10^{-1}}\right)
 \left(\frac{1 \rm{TeV}}{m_{\phi}}\right)^{1/2}
 \left(\frac{T_R}{1 \rm{TeV}}\right)\left(\frac{M^3}{\lambda 
M_P^3}\right)^{1/2},
\end{eqnarray}
where $N_Q$ is the number of produced NLSP particles per one leptonic
charge.

Such produced NLSPs decay into axino LSP with a typical lifetime of
$\cal{O}$$(0.1-1)$ second. Thus, NLSPs produced by the $Q$-ball decay
become a source of axino production. Of course, like gravitinos, axinos
can also be produced by other processes such as thermal processes (TP),
namely the scatterings and decays in the thermal bath, and non-thermal
processes (NTP), say the late decay of NLSPs produced thermally. The
relevant Boltzmann equations can be written as
\begin{eqnarray}
&& \dot{n}_{\rm NLSP}+3Hn_{\rm NLSP} =
 -\langle \sigma v\rangle (n_{\rm NLSP}^2 - n^{{\rm eq}}_{\rm NLSP}{}^2)
 +\gamma_{\rm Q-ball}-\Gamma_{\rm NLSP}n_{\rm NLSP}, \\
&& \dot{n}_{\tilde{a}}+3Hn_{\tilde{a}} =
 \langle\sigma v(i+j \rightarrow \tilde{a}+ ...)\rangle{}_{ij} n_i n_j
 + \langle\sigma v(i \rightarrow \tilde{a}+ ...)\rangle{}_{i} n_i 
 +\Gamma_{\rm NLSP}n_{\rm NLSP} ,
\end{eqnarray}
where $\gamma_{\rm Q-ball}$ denotes the contribution to NLSP production
by $Q$-balls decay, $\langle\sigma v\rangle{}_{ij}$ and $\langle\sigma
v\rangle{}_i$ are the scattering cross section and the decay rate for
the thermal production of axinos, and $\Gamma_{\rm NLSP}$ is the decay rate
of the NLSP. The total NLSP abundance, before its decay, is given by
\begin{equation}
Y_{\rm NLSP} = N_Q \frac{n_L}{s} + Y_{\rm NLSP}^{\rm TP},
\end{equation}
where $N_Q n_L/s$ denotes the NLSP produced by $L$-ball decay and
 $Y_{\rm NLSP}^{\rm TP}$ is the abundance of NLSP produced thermally and given by
\begin{equation}
 Y_{\rm NLSP}^{\rm TP} \simeq \left.\frac{H}{s}\right|_{T=m_{\rm NLSP}}
  \frac{m_{\rm NLSP}/T_f}{\langle\sigma v\rangle_{ann}} .
\end{equation}
Here $\langle\sigma v\rangle_{ann}$ is the annihilation cross section
and $T_f \sim m_{\rm NLSP} /20$ is the freeze-out temperature. The
resultant total axino abundance is expressed as
\begin{equation}
Y_{\tilde{a}} = Y_{\tilde{a}}^{\rm NTP}+ Y_{\tilde{a}}^{\rm TP} .
\end{equation}
Here
\begin{equation}
Y_{\tilde{a}}^{\rm NTP} =Y_{\rm NLSP}
 = N_Q \frac{n_L}{s} + Y_{\rm NLSP}^{\rm TP} 
\end{equation}
is the nonthermally produced axino through the NLSP decay and
$Y_{\tilde{a}}^{\rm TP}$ denotes the axinos produced by thermal
processes. For nonthermally produced axinos, 
while the NLSP abundance produced by $L$-ball decay is
\begin{equation}
N_Q \frac{n_L}{s} = 2 \times 10^{-9} \left(\frac{N_Q}{1}\right)
 \left(\frac{n_L/s}{2\times 10^{-9}}\right) ,
\end{equation}
the typical value of $Y_{\rm NLSP}^{\rm TP}$ is given by
\beq
  Y_{\rm NLSP}^{\rm TP} \simeq 10^{-11} 
       \lmk \frac{100 {\rm GeV}}{m_{\rm NLSP}} \rmk 
       \lmk \frac{10^{-10} {\rm GeV^{-2}}}{\langle\sigma v\rangle_{ann}} \rmk. 
\eeq
Thus, nonthermal production of axinos due to the thermal relic NLSPs
decay, $Y_{\rm NLSP}^{\rm TP}$, can be negligible compared to that 
from $Q$-ball produced NLSPs, $N_Q n_L/s$. 
On the other hand, axino production by thermal processes is
dominated by scattering processes for the case that the reheating
temperature is larger than the masses of neutralinos and gluinos. In
this case, the abundance of such axinos is proportional to the reheating
temperature $T_R$ and the inverse square of Peccei-Quinn (PQ) scale $f_a$ 
and given by
\cite{ckkr}
\begin{equation}
Y_{\tilde{a}}^{\rm TP}
 \simeq 10^{-8}\left(\frac{T_R}{1 {\rm TeV}}\right)
 \left(\frac{10^{11} {\rm GeV}}{f_a/N}\right)^2 ,
\end{equation}
where $N$ is the number of vacua and $N=1(6)$ for the KSVZ (DFSZ) model
\cite{KSVZ,DFSZ}. Thus, for $T_R \simeq 1$ TeV, if $f_a/N \gtrsim$
several $\times 10^{11}$ GeV \cite{YKY}, $Y_{\tilde{a}}^{\rm TP}$ is
subdominant compared with $Y_{\tilde{a}}^{\rm NTP} \simeq N_Q n_L/s$.

If this is the case, the energy density of axino is given by
$\rho_{\tilde{a}} = m_{\tilde{a}} n_{\rm NLSP}$ due to the $R$-parity
conservation.  Recalling
\begin{equation}
\frac{\rho_{DM}}{s}
 \simeq 3.9 \times 10^{-10}\left(\frac{\Omega_{DM} h^2}{0.11}\right) \rm{GeV},
\end{equation}
the density parameter of axinos is expressed as
\begin{equation}
\frac{\Omega_{\tilde{a}} h^2}{0.11}
 \simeq \left(\frac{m_{\tilde{a}}}{0.2 \rm{GeV}}\right)
 \left(\frac{n_L/s}{2\times 10^{-9}}\right) 
 \left(\frac{N_Q}{1}\right). 
 \label{OmegaAxino}
\end{equation}
Thus, axinos with the sub-GeV mass can be dark matter in our
scenario. Now, one can see that the $\Omega_b$ and $\Omega_{DM}$ is
related through the lepton asymmetry. In fact, from
Eqs.~(\ref{BaryonAsymmetry}) and (\ref{OmegaAxino}), we obtain a
relation between the abundances of dark matter and baryon asymmetry,
\begin{equation}
\frac{\Omega_b}{\Omega_{\tilde{a}}}
 \simeq \frac{2}{11}\left(\frac{N_Q}{1}\right)
 \left(\frac{\Delta Q/Q}{1 \times 10^{-1}}\right)
 \left(\frac{0.2}{m_{\tilde{a}}/m_{\rm p}}\right) ,
\label{Omegab-OmegaDM}
\end{equation}
where $m_{\rm p} (\simeq 1$ GeV) is the mass of proton.  One may find
the similar relation in the case of baryonic $Q$-ball
($B$-ball)~\cite{RoszSeto}.  The difference between the case of $B$-ball
and $L$-ball is that the required mass of LSP from $L$-ball can be an
order of magnitude smaller than that in $B$-ball where the mass of LSP
dark mater must be $\simeq 1$ GeV, mainly because a part of lepton
asymmetry produced by the Affleck-Dine mechanism, that is, only
evaporated charges $\Delta Q/Q$ are converted to baryon asymmetry so
that the number density of NLSPs produced by the $L$-ball decay become
larger for a fixed baryon asymmetry. As shown in the next section, such
a difference of axino masses is crucial for solving the missing
satellite problem and the cusp problem.

Equation~(\ref{OmegaAxino}) with $T_R \simeq 1$ TeV to explain the
observed baryon asymmetry yields a quite natural scale of the AD field
mass
\begin{equation}
m_{\phi} \simeq  1 {\rm TeV}
 \left(\frac{\epsilon}{4 \times 10^{-1}}\right)^2
 \left(\frac{M^3}{\lambda M_P^3}\right) 
 \left(\frac{m_{\tilde{a}}}{0.2 \rm{GeV}}\right)^2.
\end{equation}
As mentioned above, in our scenario, NLSP decays into axino at late
time.  Such late decay is potentially constrained by BBN.  The lifetime
of NLSP is given as
\begin{equation}
\tau_{\chi} \equiv \tau(\chi \rightarrow \tilde{a}+\gamma) =
 0.33 {\sec} \frac{1}{C_{aYY}^2Z_{11}^2}
 \left(\frac{\alpha_{em}}{1/128}\right)^{-2}
 \left(\frac{f_a/N}{10^{11}{\rm GeV}}\right)^2
 \left(\frac{10^2 {\rm GeV}}{m_{\chi}}\right)^3
\label{NLSPlifetime}
\end{equation}
for the case that the lightest neutralino $\chi$ is NLSP in~\cite{ckkr}.  Here,
$C_{aYY}$ is the axion model dependent coupling coefficient between
axion multiplet and $U(1)_Y$ gauge field, $Z_{11}$ denotes the fraction
of $b$-ino component in the lightest neutralino.  According to
Ref.~\cite{ckkr}, we can summerize the constraints as follows. First of
all, for $\tau_{\chi} \leq 0.1$ sec., there is no constraint. The
corresponding mass of the NLSP neutralino is
\begin{equation}
m_{\chi}
 = 320 {\rm GeV} \left(\frac{0.1 {\sec}}{\tau_{\chi}}\right)^{1/3}
 \left(\frac{1}{C_{aYY}^2Z_{11}^2}\right)^{1/3}
 \left(\frac{f_a/N}{\sqrt{10}\times 10^{11}{\rm GeV}}\right)^{2/3}
\end{equation}
from Eq.~(\ref{NLSPlifetime}).  Thus, if $m_{\chi} \gtrsim 320$ GeV,
this model is free from problems by the late decay of NLSP. This lower
bound is a bit stringent than that in \cite{ckkr}, because we need to
take the PQ scale somewhat larger as we mentioned.  On the other hand,
for $0.1$ sec. $< \tau_{\chi} < 1$ sec., the lower
bound on axino mass exists and can be roughly expressed as
\begin{eqnarray}
&& \frac{m_{\tilde{a}}}{0.1 {\rm GeV}} \gtrsim
 -4 \left(\frac{m_{\chi}}{10^2 {\rm GeV}}\right) (C_{aYY}Z_{11})^{2/3}
 \left(\frac{f_a/N}{10^{11}{\rm GeV}}\right)^{-2/3} + 6 
  \simeq -4 \left(\frac{0.33 {\rm sec.}}{\tau_{\chi}}\right)^{1/3} + 6, 
 \nonumber \\
 && 
 \label{NLSPconst}
\end{eqnarray}
by reading Fig. 4 in Ref.~\cite{ckkr}.
In this case, axino must be heavier than a few hundred MeV.  For
$\tau_{\chi} \simeq 1$ sec., the corresponding mass of the NLSP
neutralino and the lower bound of axino mass is given by
\begin{eqnarray}
&& m_{\chi} \simeq 150 {\rm GeV}, \nonumber \\
&& m_{\tilde{a}} \gtrsim 320 {\rm MeV}.
\end{eqnarray}

\section{A solution to the missing satellite problem and the cusp problem}

An interesting consequence of such light axinos is their large velocity
dispersion. Therefore, they can potentially solve the missing satellite
problem and the cusp problem as stated in the introduction.  For the
scenario of dark matter particle produced by the late decay of
long-lived particle, it is shown that the missing satellite problem and
the cusp problem can be solved simultaneously if the lifetime of
long-lived particle and the ratio of mass between dark matter particle
and the mother particle satisfy the following relation
\cite{HisanoInoueTomo}:
\begin{eqnarray}
  \lmk \frac{ 6.3 \times 10^2\,m_{\tilde{a}} }{m_{\chi}} \rmk^2 
    {\rm sec.} 
  \lesssim \tau_{\chi} \lesssim
  \lmk \frac{ 1.0 \times 10^3\,m_{\tilde{a}} }{m_{\chi}} \rmk^2
    {\rm sec.}, 
 \label{Lifetime-MassRatio}
\end{eqnarray}
where we identified dark matter with axino (LSP) and the long-lived
particle with the lightest neutralino (NLSP), respectively.

Combining Eq. (\ref{NLSPlifetime}) with Eq. (\ref{Lifetime-MassRatio}), we
have the following relation between the masses of the axino (LSP) and
the lightest neutralino (NLSP),
\bea
  m_{\chi} \simeq (0.33 - 0.83) 
                   \lmk \frac{m_{\tilde{a}}}{1 {\rm GeV}} \rmk^{-2}
                   \left(\frac{1}{C_{aYY}^2Z_{11}^2}\right)
                   \left(\frac{\alpha_{em}}{1/128}\right)^{-2}
                   \left(\frac{f_a/N}{10^{11}{\rm GeV}}\right)^2
                   {\rm GeV}.
\label{massrelation}
\eea
For $m_{\tilde{a}} \simeq 1$ GeV in the $B$-ball case \cite{RoszSeto},
$m_{\chi}$ becomes a few GeV with its lifetime $\tau_{\chi} \gg 10^3$
second even if we take $f_a/N$ to be several times $10^{11}$ GeV, which
is excluded. On the other hand, for $m_{\tilde{a}} = \CO(0.1)$ GeV in
the $L$-ball case of this paper, $m_{\chi}$ becomes $\CO(100)$ GeV with
its lifetime $\tau_{\chi} \lesssim 1$ second if we take $f_a/N$ to be
several times $10^{11}$ GeV. Thus, we find that axinos in this scenario
can solve the missing satellite problem and the cusp problem
simultaneously for natural mass scales with $m_{\tilde{a}} = \CO(0.1)$
GeV and $m_{\chi} = \CO(100)$ GeV.

\section{Concluding remarks}

In this paper, we show that Affleck-Dine leptogenesis can explain baryon
asymmetry and dark matter abundance simultaneously and that $\Omega_b -
\Omega_{DM}$ coincidence is explained for sub-GeV mass of the LSP axino.
Though the basic idea is the same as Ref.~\cite{RoszSeto}, where
$B$-balls are considered, the mass of LSP axino becomes an order of
magnitude smaller in this scenario. On the other hand, the PQ scale is
determined as $f_a/N =$ a few $\times 10^{11}$ GeV, which will be tested
in the future if PQ scale can be measured by e.g., the manner proposed
in~\cite{bchrs}.

The other attractive point is that axinos considered in this paper can
potentially solve the missing satellite problem and the cusp problem
simultaneously because they are relatively light and have large velocity
dispersion. We have shown that axions in our scenario can solve both
problems for natural mass scales with $m_{\tilde{a}} = \CO(0.1)$ GeV and
$m_{\rm NLSP} = \CO(100)$ GeV.  For simplicity, we concentrate on the case
of the lightest neutralino NLSP with the mass to be $\CO(100)$ GeV. Such
neutralinos are detectable in Large Hadron Collider. 
In addition, the corresponding lepton
asymmetry is almost the maximal value under the assumption of
$(M^3/\lambda M_P^3)\simeq 1$.  Hence, if this model is the simultaneous
answer to both $\Omega_b - \Omega_{DM}$ coincidence, the missing
satellite problem, and the cusp problem, we would discover the NLSP
neutralinos \footnote{Of course, NLSP might be another kind of particle,
although we do not discuss here.} with the mass of $\CO(100)$ GeV and
their decays into photons (and negligible missing energy) with the
lifetime of about $(0.1-1)$ second.

%
\section*{Acknowledgments}

We thank M. Kawasaki and T. Takahashi for useful comments. 
We are grateful to John McDonald for pointing out an error in 
the power of a coefficient in an earlier version.
O.S. would like to thank RESCEU at the University of Tokyo for the hospitality
where this work was initiated. 
The work of O.S. is supported by the MEC project FPA 2004-02015 and 
the Comunidad de Madrid project HEPHACOS (No.~P-ESP-00346).
M.Y. is supported in part by JSPS
Grant-in-Aid for Scientific Research No.~18740157 and the project of the
Research Institute of Aoyama Gakuin University.




\begin{thebibliography}{99}

\bib{AD}
I. Affleck and M. Dine,
\NPB{249}{361}{85}.

\bib{coleman}
S. Coleman,
\NPB{262}{263}{85}.

\bibitem{KasuyaKawasaki}
S.~Kasuya and M.~Kawasaki, Phys.\ Rev.\ D {\bf 62}, 023512 (2000).

\bib{KS}
A. Kusenko and M. Shaposhnikov,
\PLB{418}{46}{98}.

\bibitem{QballEnqvist} 
K.~Enqvist and J.~McDonald, Phys.\ Lett.\ B {\bf 425}, 309 (1998);
K.~Enqvist and J.~McDonald, Nucl.\ Phys.\ B {\bf 538}, 321 (1999).

\bib{EM2}
K. Enqvist and J. McDonald,
\PLB{440}{59}{98};
\NPBB{570}{407}{00}.

\bibitem{FHY}
M.~Fujii, K.~Hamaguchi and T.~Yanagida, Phys.\ Rev.\  D {\bf 64}, 123526 (2001).

\bibitem{FH}
M.~Fujii and K.~Hamaguchi, Phys.\ Lett.\  B {\bf 525}, 143 (2002); 
Phys.\ Rev.\  D {\bf 66}, 083501 (2002).

\bibitem{Seto}
O.~Seto, Phys.\ Rev.\ D {\bf 73}, 043509 (2006).

\bib{BBP}
M. Bolz, W. Buchmuller, and M. Plumacher,
\PLB{443}{209}{98}.

\bibitem{Pradler}
For a recent anlysis, see e.g., 
J.~Pradler and F.~D.~Steffen, Phys.\ Rev.\  D {\bf 75}, 023509 (2007);
Phys.\ Lett.\  B {\bf 648}, 224 (2007). 

\bib{FRT}
J. L. Feng, A. Rajaraman, and F. Takayama,
\PRLL{91}{011302}{03}.

\bib{FST}
J. L. Feng, S. Su, and F. Takayama,
\PRDD{70}{075019}{04}.

\bibitem{RoszSeto}
L.~Roszkowski and O.~Seto, Phys.\ Rev.\ Lett.\  {\bf 98}, 161304 (2007).

\bib{RTW}
K. Rajagopal, M. S. Turner, and F. Wilczek,
\NPB{358}{447}{91}.

\bib{KMN}
J. E. Kim, A. Masiero, and D. V. Nanopoulos,
\PLBold{139}{346}{84}.

\bib{CKR}
L. Covi, J. E. Kim, and L. Roszkowski,
\PRL{82}{4180}{99}.

\bibitem{ckkr}
L.~Covi, H.~B.~Kim, J.~E.~Kim and L.~Roszkowski, 
\JHEPP{0105}{033}{01}.

\bibitem{Brandenburg}
A.~Brandenburg and F.~D.~Steffen, JCAP {\bf 0408}, 008 (2004).

\bib{MSP}
A. A. Klypin, A. V. Kravtsov, O. Valenzuela, and F. Prada,
\APJ{522}{82}{99};
B. Moore, {\it et al.},
\APJ{524}{L19}{99};
A. R. Zentner and J. S. Bullock,
\APJJ{598}{49}{03}.

\bib{CP}
B. Moore,
\NAT{370}{629}{94};
R. A. Flores and J. A. Primack,
\APJ{427}{L1}{94};
W. J. G. De Blok and S. S. McGaugh,
\MNRAS{290}{533}{97};
J. J. Binney and N. W. Evans,
\MNRASS{327}{L27}{01};
A. R. Zentner and J. S. Bullock,
\PRDD{66}{043003}{02};
J. D. Simon {\it et al.},
\APJJ{621}{757}{05}.

\bib{BKW}
J. S. Bullock, A. V. Kravtsov, and D. H. Weinberg,
\APJJ{539}{517}{00}.

\bibitem{Yokoyama}
For example, see J.~Yokoyama, 
Phys.\ Rev.\  D {\bf 62}, 123509 (2000).
  
\bib{SS}
D. N. Spergel and P. J. Steinhardt,
\PRLL{84}{3760}{00}.

\bib{WDM}  
P. Colin, V. Avila-Reese, and O. Valenzuela,
\APJJ{542}{622}{00};
J.~Hisano, K.~Kohri and M.~M.~Nojiri, Phys.\ Lett.\  B {\bf 505}, 169 (2001);
W.~B.~Lin, D.~H.~Huang, X.~Zhang and R.~H.~Brandenberger,
  Phys.\ Rev.\ Lett.\  {\bf 86}, 954 (2001).
P. Bode, J. P. Ostriker, and N. Turok,
\APJJ{556}{93}{01};
J. L. Feng, A. Rajaraman, and F. Takayama,
\PRLL{91}{011302}{03};
J. A. R. Cembranos, J. L. Feng, A. Rajaraman, and F. Takayama,
\PRLL{95}{181301}{05};
M.~Kaplinghat, Phys.\ Rev.\  D {\bf 72}, 063510 (2005).

\bibitem{Ng}
K.~W.~Ng, Nucl.\ Phys.\ B {\bf 321}, 528 (1989).

\bibitem{DineRandallThomas}
M.~Dine, L.~Randall and S.~Thomas, Nucl.\ Phys.\ B {\bf 458}, 291 (1996).

\bibitem{K}
K.~Enqvist and J.~McDonald, Phys.\ Lett.\ B {\bf 425}, 309 (1998); 
K.~Enqvist, A.~Jokinen and J.~McDonald, Phys.\ Lett.\ B {\bf 483}, 191 (2000).

\bibitem{ThermalMass}
R.~Allahverdi, B.~A.~Campbell and J.~R.~Ellis,
Nucl.\ Phys.\ B {\bf 579}, 355 (2000).

\bibitem{TwoLoop}
A.~Anisimov and M.~Dine, Nucl.\ Phys.\ B {\bf 619}, 729 (2001).

\bibitem{GravitinoProblem}
M.~Y.~Khlopov and A.~D.~Linde, 
Phys.\ Lett.\ B {\bf 138}, 265 (1984); 
J.~R.~Ellis, J.~E.~Kim and D.~V.~Nanopoulos, 
Phys.\ Lett.\ B {\bf 145}, 181 (1984).

\bibitem{BanerjeeJedamzik}
R.~Banerjee and K.~Jedamzik, 
Phys.\ Lett.\ B {\bf 484}, 278 (2000).

\bibitem{Joyce:1994zn}
M.~Joyce, T.~Prokopec and N.~Turok, 
Phys.\ Rev.\  D {\bf 53}, 2930 (1996).

\bibitem{DW}
H. Davoudiasl and E. Westphal,
\PLB{432}{128}{98}.

\bib{sphaleron}
V. A. Kuzmin, V. A. Rubakov, and M. E. Shaposhnikov,
\PLBold{155}{36}{85}; 
S. Y. Khlebnikov and M. E. Shaposhnikov,
\NPB{308}{885}{88}; 
J. A. Harvey and M. S. Turner,
\PRD{42}{3344}{90}.

\bib{KSVZ}
J. E. Kim, 
\PRL{43}{103}{79};
M. A. Shifman, V. I. Vainshtein, and V. I. Zakharov,
\NPB{166}{493}{80}.

\bib{DFSZ}
M. Dine, W. Fischler, M. Sredicki,
\PLBold{104}{199}{81};
A. R. Zhitnitsky, Yad. Fiz. {\bf 31}, 497 (1980) [Sov. J. Nucl. Phys. {\bf 31}, 260 (1980)].

\bib{YKY}
For the latest constraint on the breaking scale of the PQ symmetry, see
M. Yamaguchi, M. Kawasaki, and J. Yokoyama,
\PRL{82}{4578}{99}.

\bibitem{HisanoInoueTomo}
J.~Hisano, K.~T.~Inoue and T.~Takahashi, 
Phys.\ Lett.\  B {\bf 643}, 141 (2006).

\bibitem{bchrs}
A.~Brandenburg, L.~Covi, K.~Hamaguchi, L.~Roszkowski and F.~D.~Steffen,
Phys.\ Lett.\  B {\bf 617}, 99 (2005).


\end{thebibliography}
\end{document}